\documentclass[sn-mathphys,iicol]{sn-jnl}

\jyear{2022}%

\theoremstyle{thmstyleone}%
\theoremstyle{thmstyletwo}%

\theoremstyle{thmstylethree}%

\usepackage{amsthm}
\begin{document}

\title[Article Title]{Effect of the interference in overlapped double-pulse irradiation at the silicon surface}


\author*[1,2]{\fnm{Tomohito} \sur{Otobe}}\email{otobe.tomohito@qst.go.jp}

\author[1]{\fnm{Prachi} \sur{Venkat}}\email{venkat.prachi@qst.go.jp}

\affil*[1]{\orgdiv{KPSI}, \orgname{National Institutes for Quantum Science and Technology}, \orgaddress{\street{8-1-7, Umemidai}, \city{Kizugawa}, \postcode{619-0125}, \state{Kyoto}, \country{Japan}}}
\affil[2]{\orgdiv{Photon Science Center}, \orgname{The University of Tokyo}, \orgaddress{\street{7-3-1, Hongo}, \city{Bunkyo-ku}, \postcode{111-8656}, \state{Tokyo}, \country{Japan}}}


\abstract{We studied the excitation process of silicon under an intense double pulse. We employed the three-temperature (electron, hole, and lattice) model
(3TM) together with  Maxwell's equations. We solved Maxwell's equations by the finite-difference time-domain approach.
The lattice temperature and absorbed energy at the surface increase significantly when the two laser pulses overlap with 
constructive interference. On the other hand, destructive interference reduces the efficiency of laser excitation significantly. 
On an average, the overlapped double pulse increases the efficiency to about twice of the distinct two-pulse case.}




\maketitle

\section{Introduction}\label{sec1}
In the recent years, we have been able to use intense laser pulses of duration ranging from femto- to pico-second time scales to investigate laser-matter interaction. In particular, laser processing of semiconductors using ultrashort pulses attracts great interest, due to its application in nano-structuring \citep{Gattass2008,Stoian-2020}. 
Ultrafast laser pulse enables precise processing of the material, without causing damage in the surrounding area \citep{Sugioka2014}.
There are two approaches to study the laser processing, single-shot and multi-shot. 

The single-shot experiment is suitable to study excitation process during the laser pulse\cite{Kumada-2014,Kumada-2016,Dinh-2019}, while the multi-shot experiments can study how to control the efficiency of laser processing\cite{Nakamura-2020,Otobe-2017}.
In the multi-shot experiment, a pulse-train with distinct laser pulses  is employed. Recently, enhancement of laser ablation and processing by an overlapped double-pulse irradiation has been reported \cite{Zhang_2015,Zhang_2019,Wang-2020}. 

\begin{figure}%
\centering
\includegraphics[width=0.45\textwidth]{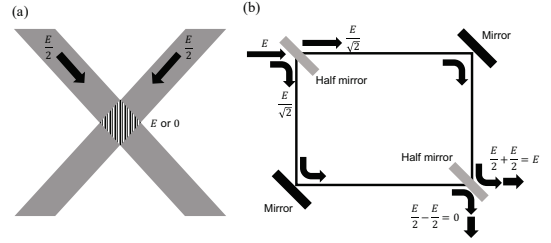}
\caption{Tow possible way to make double pulse. (a) Non-collinear configuration by the two sources. 
(b) Collinear configulation by the interferometer. }\label{fig1}
\end{figure}

When the two pulses are overlapped, the interference between them
plays a crucial role as shown in Fig.~\ref{fig1}.
If the two pulses with same irradiance ($Ir\propto E^2$) overlap perfectly, constructive interference makes 4-times intense laser field ($4 Ir\propto (2E)^2$), 
while the destructive interference makes a zero intensity region\cite{Drosd-2014}.
In the case of non-collinear configuration (Fig.~\ref{fig1}(a)), 
position-dependent intensity distribution occurs.
The collinear configuration (Fig.~\ref{fig1}(b)) makes constructive and destruction pulse after the second half mirror. 
Since the laser intensity at the target depends on the relative 
phase of two pulses, overlapped double-pulse experiments have been done between
$4 Ir$ and $0$ intensity,
we should take into account the effect of 
interference. 

Some theoretical approaches have been proposed as the extension of 
two-temperature-model (TTM)\cite{Anisimov1974,Chen-2005,Silaeva_2012}.
Recently, we have developed a new numerical model 
including the time-evolution of electron, hole, and lattice temperatures (3 temperature model=3TM),
for the laser excitation process of the silicon \cite{Prachi-2022}. 
One of the features of 3TM is the direct treatment 
of electromagnetic field (EM-field) 
with finite-difference time-domain (FDTD) method.
In the previous work, we have reported that the 3TM-FDTD reproduces
the pulse-duration dependence of the damage threshold of silicon
not only qualitatively but also quantitatively.


This paper is organized as follows: Section \ref{form} describes the analytical formulation of 3TM and numerical method for FDTD method, section \ref{results} presents results of excitation in silicon for different delay time of the two pulses and laser intensity, and then finally, we would like to summarize the work in section \ref{summary}.

\section{Formulation}\label{form}
\subsection{3TM}
The time-evolution of electron and hole densities, $n_e$ and $n_h$ is described as:
\begin{align}
\label{source}
        \frac{\partial n_{e(h)}}{\partial t} & = \frac{\alpha I}{\hbar \omega_0} + \frac{\beta I^2}{2\hbar \omega_0} \nonumber\\
        &- \gamma_e n_en_en_h -\gamma_h n_hn_hn_e \nonumber\\
     & + \frac{1}{2}(\theta_e n_e + \theta_h n_h)  \nonumber\\
     &+ \nabla D_{e(h)}\cdot\vec J_{e(h)}+ D_{e(h)}\nabla\cdot \vec J_{e(h)} \nonumber\\
     & -(+) \mu_{e(h)}(\nabla\cdot n_{e(h)}\vec{F} +\nabla n_{e(h)}\cdot\vec{F}\nonumber\\ 
     &  +n_{e(h)} \nabla\cdot \vec{F})
\end{align}
where $\omega_0$ is the laser frequency and $\alpha$ is the single photon absorption coefficient for transition from VB to CB \citep{Green2008}. $\beta$ is the two-photon absorption coefficient for which we use the interpolation of the DFT calculation when $2\hbar\omega_0 > E_o$ \cite{Murayama-1995}, where $E_o$ is the optical gap, and  we employ the model described in Ref.\cite{Garcia_2006,Furey-2021} for $E_o \geq 2\hbar\omega_0 \geq E_g$. $\gamma_{e(h)}$ is the Auger re-combination coefficient \citep{Silaeva_2012} and $\theta_{e(h)}$ is the impact ionization coefficient \citep{Chen-2005}. Equation (\ref{source}) also includes the effect of spatial charge distribution and the associated electric field, and $J_{e(h)}, D_{e(h)}$ and $\vec{F}$ are the charge current,
\begin{align}
        &\vec J_{e(h)}=\nabla n_{e(h)} +\frac{m_{r,{e(h)}}n_{e(h)}H_{1/2}^{-1/2}(\eta_{e(h)})}{ k_B T_{e(h)}}\nabla E_g\nonumber\\ &+\frac{n_{e(h)}}{T_{e(h)}}\left(2H_{1/2}^{-1/2}(\eta_{e(h)})H_{0}^{1}(\eta_{e(h)}) -\frac{3}{2}\right)\nabla T_{e(h)}    
\end{align}
diffusion coefficient
\begin{equation}
    D_{e(h)}=\frac{\mu^0_{e(h)}k_B T_{e(h)}}{-(+)e} H_{-1/2}^{0}(\eta_{e(h)}),
\end{equation}
 and the electric field induced by the electron--hole separation \cite{Silaeva_2012}, respectively.
 Here $\eta_{e(h)}$ is the reduced Fermi level:
\begin{equation}
\label{eq:eta_e}
    \eta_e = \frac{\phi_e - E_C}{k_BT_e}   
\end{equation}
\begin{equation}
\label{eq:eta_h}
    \eta_h = \frac{E_V-\phi_h}{k_BT_h},   
\end{equation}
and
\begin{equation}
  H_{\xi}^{\zeta}(\eta)=\frac{F_{\zeta}(\eta)}{F_{\xi}(\eta)}   
\end{equation}
where $F_{\xi}(\eta)$ is the Fermi integral given by 
\begin{equation}
    F_{\xi}(\eta) = \frac{1}{\Gamma(\xi+1)}\int_{0}^{\infty}dx\frac{x^\xi}{1+\exp(x-\eta)}.
\end{equation}
$F_{\xi}(\eta)$ are used depending on whether $\eta$ is positive \cite{FUKUSHIMA2015802} or negative \cite{Cloutman-1989}.

The total dielectric function along with the effect of band structure re-normalization \citep{Sokol-2000} is expressed by 
\begin{equation}
    \epsilon(\omega) = 1 + \frac{n_0 - n_e}{n_0}\epsilon_L(\omega + \delta E_g/\hbar) + \epsilon_D(\omega) 
\end{equation}
where $n_0$ is the density of valence electrons. It should be noted that $n_e$ and $n_h$ are nearly the same due to the effect of $\vec{F}$ and can be approximated at $n_e$. $\epsilon_L(\omega)$ is the innate dielectric function, $\delta E_g$ represents the band re-normalization by carrier density, and $\epsilon_D$ is the complex  dielectric function calculated from Drude model.
The temperature dependent optical parameters of silicon are referred to from Ref.\citep{Green2008}. 
Since this temperature dependence in $\epsilon_L(\omega)$ does not 
include band re-normalization, we shift the photon energy by $\delta E_g$ (Tab.~\ref{table2}).
$\epsilon_D$  accounts for the effect of plasma in the excited system. 
Considering the electron and hole sub-systems,
\begin{align}
        \epsilon_D(\omega) =  -\frac{4\pi n_e e^2}{\omega_0^2}&\Bigg[ \frac{1}{m^*_{e,cd}\left(1 + i\frac{\nu_e}{\omega_0}\right)} \nonumber\\
        &+ \frac{1}{m^*_{h,cd}\left(1 + i\frac{\nu_h}{\omega_0}\right)} \Bigg]
\end{align}
where $m^*_{e(h),cd}$ is the effective mass for the conductivity \cite{Ramer-2014} and  $\nu_e$ and $\nu_h$ are the collision frequencies, describing the electron-hole ($e-h$), electron-phonon ($e-ph$) and hole-phonon ($h-ph$) collisions. The $e-ph$ and $h-ph$ collisions are assumed to have the same frequency which is dependent on lattice temperature \citep{Ramer-2014}. Effect of electron-ion core collisions is also considered \cite{Sato-2014}. 
The electron temperature dependence of the electron-ion core collisions is fitted to the damping time data from Sato \textit{et.al.} \cite{Sato-2014} as
$
\tau_{e-ion}=0.98+0.2(k_B T)^{-3.5}~\textrm{fs}.
\label{e-ionfit}
$
The collision frequencies for $e-h$ interactions are calculated as per the model presented in Ref.\citep{Terashige-2015}.
The total one-photon absorption coefficient including free-carrier absorption is 
\begin{equation}
  \alpha_f=\frac{2\pi}{\lambda}\Im[\sqrt{\epsilon(\omega)}],
   \end{equation} 
where $\lambda$ is the laser wavelength in vacuum.

Since 3TM considers three sub-systems, \textit{viz.} electron, hole and lattice, their temperatures also evolve separately.
The temperature evolution is expressed as:
\begin{equation}
\begin{split}
    C_{e}&\frac{\partial T_{e}}{\partial t}=  m_{r,{e}}(\alpha_f I + \beta I^2)\\
    &+E_g\gamma_{e} n_{e} n_{e} n_{h} \\ 
    & -\frac{C_{e}}{\tau}(T_{e}-T_l)-\nabla \cdot \vec{w}_{e} \\ 
& -\frac{\partial n_{e}}{\partial t} \left(m_{r,{e}}E_g 
+ \frac{3}{2}k_B T_{e} H_{-1/2}^{1/2}(\eta_{e}) \right) \\ 
& - m_{r,{e}}n_{e}\left( \frac{\partial E_g}{\partial T_l}\frac{\partial T_l}{\partial t}
+  \frac{\partial E_g}{\partial n_{e}}\frac{\partial n_{e}}{\partial t}\right),
\end{split}
\label{Te}
\end{equation}
\begin{equation}
\begin{split}
    C_{h}&\frac{\partial T_{h}}{\partial t}=  m_{r,{h}}(\alpha_f I + \beta I^2)\\
    &+E_g\gamma_{h} n_{h} n_{h} n_{e} \\ 
    & -\frac{C_{h}}{\tau}(T_{h}-T_l)-\nabla \cdot \vec{w}_{h} \\ 
& -\frac{\partial n_{h}}{\partial t} \left(m_{r,h}E_g 
+ \frac{3}{2}k_B T_{h} H_{-1/2}^{1/2}(\eta_{h}) \right) \\ 
& - m_{r,h}n_{h}\left( \frac{\partial E_g}{\partial T_l}\frac{\partial T_l}{\partial t}
+  \frac{\partial E_g}{\partial n_{h}}\frac{\partial n_{h}}{\partial t}\right),
\end{split}
\label{Th}
\end{equation}
\begin{equation}
    C_l\frac{\partial T_l}{\partial t}=-\nabla\cdot(\kappa_l \nabla T_l)+\frac{C_e}{\tau}(T_e-T_l)+\frac{C_h}{\tau}(T_h-T_l).
    \label{Tl}
\end{equation}
The third and fourth terms in Eq.~(\ref{Te}) account for the loss of energy due to electron-lattice interaction and energy current. The last two terms on right hand side include the changes in carrier density and band gap energy.  The heat capacities $C_{e(h)}$ are defined from the internal energy as
\begin{align}
       C_{e(h)}&=\frac{\partial U_{e(h)}}{\partial T_{e(h)}}\nonumber\\ &=\frac{\partial }{\partial T_{e(h)}}n_{e(h)}\Big(m_{r,e(h)}E_g \nonumber\\
       &+\frac{3}{2} k_B T_{e(h)} H_{1/2}^{3/2}(\eta_{e(h)})\Big),
\end{align}

\begin{align}
C_e &= -\frac{3}{2}n_e k_B \eta_e \left( 1 - H_{1/2}^{3/2}(\eta_e) H_{1/2}^{-1/2}(\eta_e) \right) \nonumber\\
&+ \frac{3}{2}n_e k_B H_{1/2}^{3/2}(\eta_e)\nonumber\\
C_h &= -\frac{3}{2}n_h k_B \eta_h \left( 1 - H_{1/2}^{3/2}(\eta_h) H_{1/2}^{-1/2}(\eta_h) \right) \nonumber\\
&+ \frac{3}{2}n_h k_B H_{1/2}^{3/2}(\eta_h).
\end{align}
$\vec{w}_{e(h)}$ are the energy currents given by 
\begin{equation}
    \vec w_e = -\frac{1}{e}(2k_BT_eH_{0}^{1}(\eta_e) + m_{r,e}E_g)\vec j_e - \kappa_e\nabla T_e,
\end{equation}
and
\begin{equation}
    \vec w_h = \frac{1}{e}(2k_BT_hH_{0}^{1}(\eta_h) + m_{r,h}E_g)\vec j_h - \kappa_h\nabla T_h.
\end{equation}
Here
\begin{align}
        \vec{j_e} &=  n_e\mu_{e}^{0}H_{1/2}^{0}(\eta_e) \{ m_{r,e}\nabla E_g \nonumber\\
        &+ k_B\nabla\eta_e T_e + 2k_B H_{0}^{1}(\eta_e)\nabla T_e \}
\end{align}
and
\begin{align}
        \vec{j_h} &= -n_h\mu_{e}^{0}H_{1/2}^{0}(\eta_h) \{ m_{r,h}\nabla E_g \nonumber\\
        &+ k_B\nabla\eta_h T_h + 2k_B H_{0}^{1}(\eta_h)\nabla T_h \}
\end{align}
are the electrical current by the Zeeback effect and gradient of the 
quasi Fermi-levels of electron and hole.

\subsection{FDTD}
The propagation of the laser pulse is described by solving the Maxwell's equations using FDTD method. Mur's absorbing boundary condition is employed to prevent reflection from the boundary \citep{Mur-1981}. Another salient feature is that the electric field is considered to be complex for the calculation of laser intensity, to ensure a non-zero field at all points in time and space. Assuming a one-dimensional system, the electric field is:
\begin{equation}
    E(x,t) = E_1(x,t)e^{i\omega t}+E_2(x,t)e^{i(\omega t+\phi)}
\end{equation}
Here $E_{1(2)}$ includes Gaussian envelope, and $\phi$ is the relative phase between the two pulses. 
The electric field of the two pulses are expressed as
\begin{align}
    E_1(x,t)\vert_{x=0}&=E_0e^{-\frac{(t-T_{peak})^2}{T^2}}\nonumber,\\
    E_2(x,t)\vert_{x=0}&=E_0e^{-\frac{(t-T_{peak}-T_{d})^2}{T^2}},
\end{align}
 where $T=t_p/(4\sqrt{\ln 2})$, $t_p$ being the FWHM pulse duration, $T_{peak}$ is the time of peak intensity, and $T_d$ is the delay time.
 In this work we assume only two replica pulses.
We define the peak of the first pulse ($T_{peak}$) as $4T$ to ensure the negligible field intensity at $t=0$.
The laser irradiance ($I_r(x,t)$) is defined by the absolute value of the electric fied as
\begin{equation}
    I_r(x,t) = \frac{c}{8\pi}\Re[\sqrt{\epsilon}]\vert E(x,t)\vert^2.
\end{equation}
Evaluation of charge current induced by the laser field is a crucial part of the module. We calculate the current with and without excitation i.e., for photo-absorption and dielectric response. For dielectric response, $j_0(x,t)$ is calculated as:
\begin{equation}
    j_0(x,t) = \chi_r(\omega)\frac{\partial P(x,t)}{\partial t} = -\chi_r(\omega)\frac{\partial^2A(x,t)}{c\partial t^2}
\end{equation}
where $A(x,t)$ is the vector potential,  $P$ is the polarization and $\chi_r$ is the real part of susceptibility ($\chi = (\epsilon-1)/4\pi$).
The Maxwell's equation thus, becomes 
\begin{equation}
   \frac{1}{c^2}(1+4\pi\chi_r(\omega))\frac{\partial^2A(x,t)}{\partial t^2} - \frac{\partial^2A(x,t)}{\partial x^2} = \frac{4\pi}{c}j(x,t)
\end{equation}
where
\begin{equation}
    j(x,t) = (\alpha_f(\omega) + \beta(\omega)I_r(x,t))\frac{c\Re[\sqrt{\epsilon}]}{4\pi}E(x,t)
\end{equation}
is the current associated with photo-absorption.

\begin{sidewaystable}
        \centering
    \caption{List of symbols and notations used in the text}
    \hrule
    \vspace{0.05cm}
    \hrule
    \begin{tabular}{ l l  l l}
    $n_{e(h)}$ & electron(hole) density  & $T_{e(h)}$ & electron(hole) temperature \\
    $n_0$ & Density of valence electrons    & $T_l$ & Lattice temperature\\
    $\eta_{e(h)}$ & Reduced quasi Fermi-level & $\phi_{e(h)}$ & Quasi Fermi-level \\
    $\mu^0_{e(h)}$ & Electron(hole) mobility for Maxwell-Boltzmann distribution & 
    $E_C$ &  Lower energy limit of Conduction band\\
    $\mu_{e(h)}$ & Electron(hole) mobility &$E_V$ & Upper energy limit of Valence band\\
   $\kappa_{e(h)}$ &$\frac{k_B^2 \sigma_{e(h)} T_c}{e^2}[6H_0^2(\eta_{e(h)})-4H_0^1(\eta_{e(h)})^2]$: Thermal conductivity 
    &$m_{r,e(h)}$ & $m^*_{e(h)}/(m^*_{e}+m^*_{h})$: Ratio of the effective mass \\
    $\phi_{e(h)}$ & quasi Fermi-level & $\sigma_{e(h)}$& $-(+) e n_{e(h)}\mu_{e(h)}^0H_{1/2}^0(\eta_{e(h)})$:electrical conductivity
    \end{tabular}
    \hrule
    \vspace{0.05cm}
    \hrule
    \label{table1}
\end{sidewaystable}

\begin{sidewaystable}
    \centering
    \caption{List of parameters}
    \hrule
    \vspace{0.05cm}
    \hrule
 \begin{tabular}{ c l l  }    
    $\kappa_{l}$ & Lattice thermal conductivity \cite{VanDriel-1987} & $1585T_l^{-1.23}$ W/(cm K)  \\
    $C_l$ & Lattice heat capacity \cite{VanDriel-1987} & $1.978 + 3.54\times10^{-4}T_l-3.68T_l^{-2}$ J/(cm$^3$ K) \\
    $\tau$   & e-ph relaxation time \cite{Yoffa-1981,Sjodin-1998,Herb2006} & $\tau_0(1 + (n_e/(8\times10^{20}))^2)$ \\
    $\tau_0$ & e-ph relaxation time constant \cite{Chen-2005,Ichibayashi-2009} & 240 fs \\
    $m^{*}_{e}$ & Density-of-State effective mass of electron at 300~K  \cite{BARBER-1967,Lipp-2014} & $0.36m_e$ \\
    $m^{*}_{h}$ & Density-of-State effective mass of hole at 300~K \cite{BARBER-1967,Lipp-2014} & $0.81m_e$ \\
    $m^{*}_{e,cd}$ & Effective mass of electron for conduction \cite{Ramer-2014} & $0.26m_e$ \\
    $m^{*}_{h,cd}$ & Effective mass of hole conduction \cite{Ramer-2014} & $0.37m_e$ \\
    $\mu_e^0$ & Electron mobility \cite{Meyer-1980,Lipp-2014} & $8.5\times10^{-3}$ m$^2$/Vs \\
    $\mu_h^0$ & Hole mobility \cite{Meyer-1980,Lipp-2014} & $1.9\times10^{-3}$ m$^2$/Vs \\
    $\gamma_e$ & Auger recombination coefficient \cite{Silaeva_2012,Dziewior-1977} & $2.3\times10^{-31}$ cm$^6$/s \\
    $\gamma_h$ & Auger recombination coefficient \cite{Silaeva_2012,Dziewior-1977} & $7.8\times10^{-32}$ cm$^6$/s \\
    $\theta_{e(h)}$ & Impact ionization coefficient \cite{VanDriel-1987} & $3.6\times10^{10}\exp(-1.5E_g/k_BT_{e(h)}) s^{-1}$ \\
    $E_g$ & Band gap function & $1.16 -0.72\times 10^{-4} T_l^2/(T_l+1108) -1.5\times10^{-8}n_e^{1/3}$ eV \cite{Chen-2005}\\
    $\delta E_g$ & Band re-normalization & $-1.5\times10^{-8}n_e^{1/3}$ eV 
  \end{tabular}
    \hrule
    \vspace{0.05cm}
    \hrule
    \label{table2}
\end{sidewaystable}
\section{Results and Discussions}\label{results}



\begin{figure}%
\centering
\includegraphics[width=0.45\textwidth]{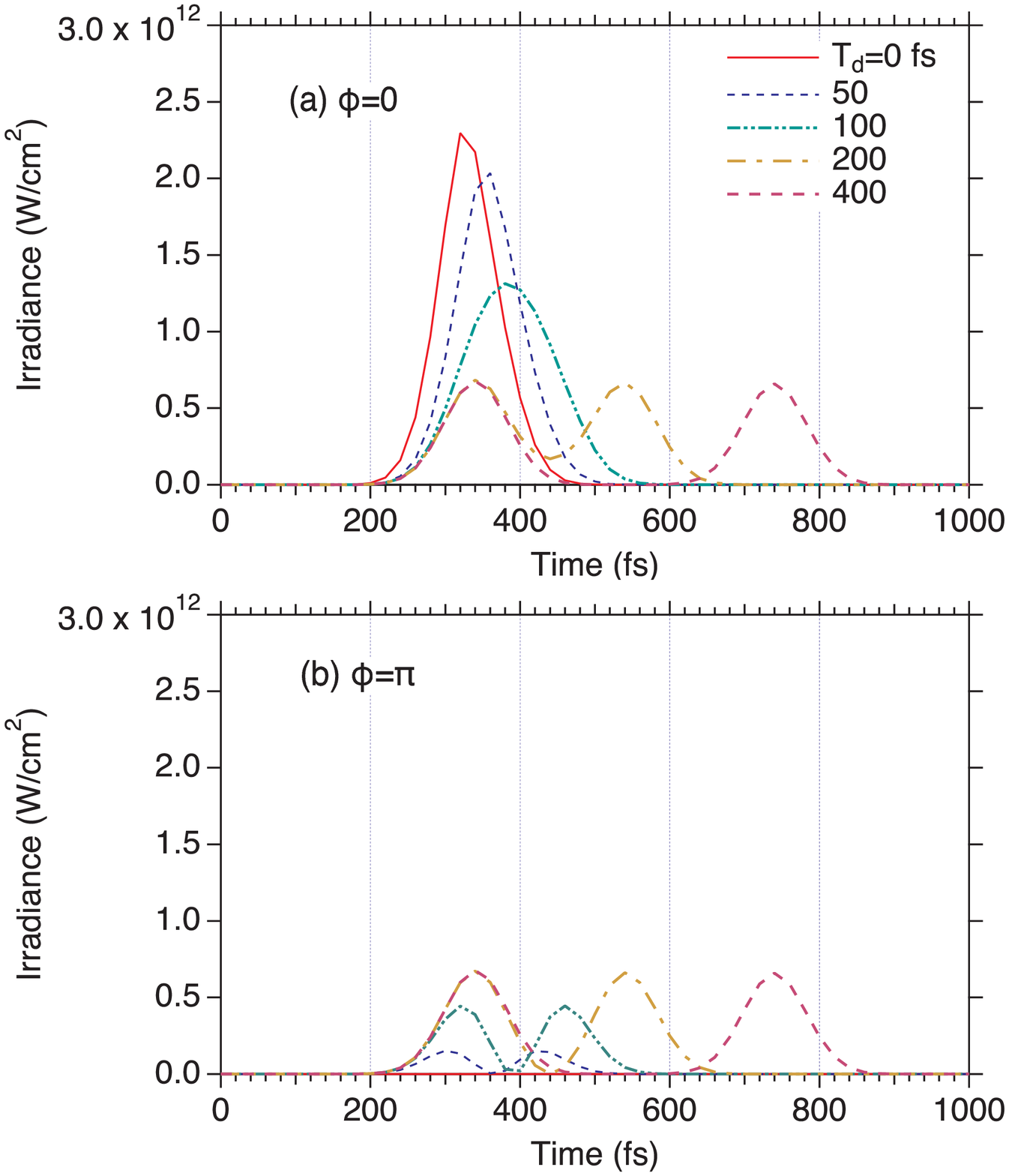}
\caption{Irradiance at the silicon surface with (a) constructive and (b) destructive interferece.
The irradiance of the incident lasers are $1\times 10^{12}$~W/cm$^2$.}\label{fig2}
\end{figure}

The behavior of the EM-field with constructive and destructive interference is the 
important point of overlapped double pulse.
Figure \ref{fig2} shows the time-evolution of laser field at the Si surface.
The irradiance of the single pulse is $1\times10^{12}$~W/cm$^2$, the frequency is 1.55~eV, and 
the pulse duration in FWHM is 100~fs.
Since our approach includes reflection at the surface, the irradiance at the surface 
is decreased from the incident pulse.

The constructive interference makes a 4-times more intense pulse with $T_d=0$~fs, while the destructive interference 
makes net zero intensity.
As $T_D$ increases, the peak intensity of pulse approaches the intensity of a single pulse.
With $T_d=200$~fs, we cannot distinguish the difference between $\phi=0$ and $\pi$.

\begin{figure}%
\centering
\includegraphics[width=0.45\textwidth]{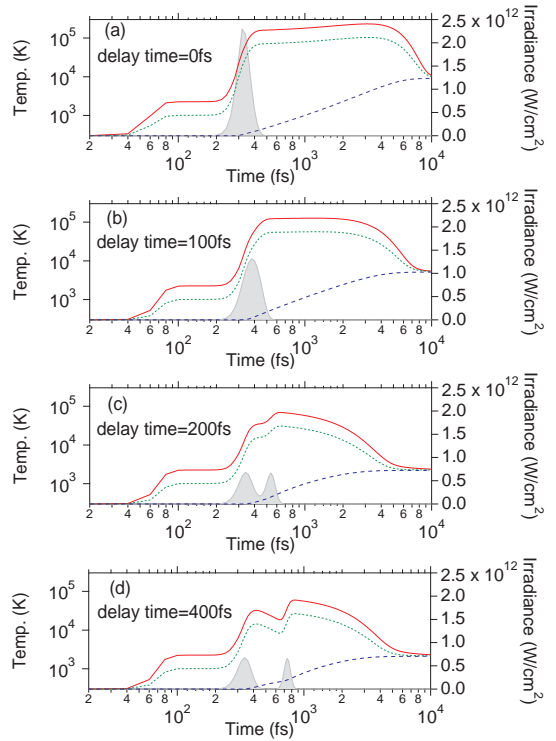}
\caption{Time-evolution of electron (red-solid), hole (green-dotted), and lattice (blue-dashed) temperatures with different
delay time.}\label{fig3}
\end{figure}

The constructive interference ($\phi=0$) may be important because it increases the efficiency of laser processing.
Figure~\ref{fig3} shows the time-evolution of electron, hole, and lattice temperatures at the surface together with the laser irradiance.
We can see an abrupt increase of electron and hole temperatures with $T_d \leq 100$~fs.
Above $T_d=200$~fs, there is a knee structure after the first pulse.
In particular, with $T_d=400$=fs, we can see a kink at the second pulse
in the $T_l$. This "kink" structure in $T_l$ may occur with $T_D$ longer than the typical time scale of electron-phonon energy transfer ($\tau_{0}=240$~fs).  

\begin{figure}%
\centering
\includegraphics[width=0.45\textwidth]{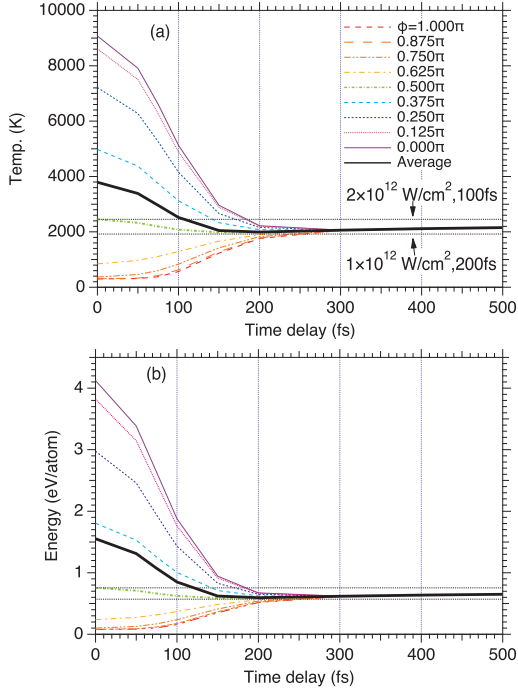}
\caption{Relative phase dependence of the (a) $T_l$ and (b) absorbed energy.}\label{fig4}
\end{figure}

In some cases, the $\phi$ is random.
The $\phi$-dependence of the $T_l$-dependence at 10~ps and absorbed energy 
are shown in Fig.~\ref{fig4}.
The $\phi$-dependence is significant in $T_d<200$~fs.
The thick-solid line indicates the average.
On average, the interference effect increases the $T_l$ to about 1.53 times and absorbed energy to about 2.5 times with respect to distinct two-pulse case ($T_d=500$~fs).

The $T_d$-dependence for $T_d>200$~fs shows slow increase as $T_d$.
This relatively weak $T_d$-dependence may due to the $T_l$-dependence
of the excitation rate.
The most intense $T_l$-dependence in our approach
is the phonon-assisted one-photon absorption in $\alpha$\cite{Green2008}. 
The increase in $\alpha$ induces the increase of $T_l$ and energy absorption
at the second pulse.

The two horizontal lines indicates the results with different laser parameters
which gives same fluence as the case of two distinct pulses.
The double-pulse excitation is more efficient than a longer  pulse (200~fs),
while it is less efficient than an intense pulse ($2\times 10^{12}$~W/cm$^2$).
Since our calculation includes the two photon absorption, higher peak intensity
induces more efficient photo-absorption.
In the case of same peak intensity, the excitation by  double pulse
 is efficient due to the $T_l$-dependence in $\alpha$.

\begin{figure}%
\centering
\includegraphics[width=0.45\textwidth]{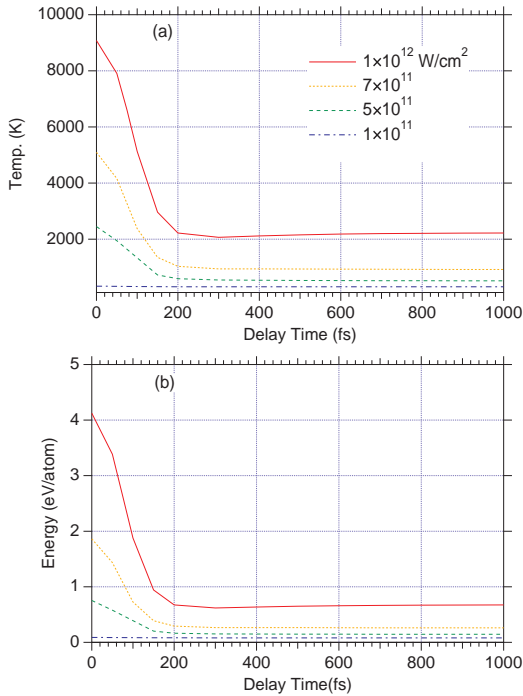}
\caption{Irradiance-dependence with constructive interference.}\label{fig5}
\end{figure}
Figure \ref{fig5} shows the incident intensity dependence of (a) $T_l$ at 10~ps and (b) absorbed energy. The intensity-dependence is significant as $T_d$ decreases. 
At intensity of $1\times10^{11}$ W/cm$^2$, the excitation is negligible with all $T_d$, while the $T_l$ exceeds melting temperature (1687~K) with $5\times10^{11}$~W/cm$^2$, and the energy exceeds the bonding energy (2.3~eV/atom) between $7\times10^{11}$ and $1\times10^{12}$~W/cm$^2$ overlapped double-pulse. 

\section{Summary}\label{summary}
In this work we apply the 3TM \cite{Prachi-2022} for silicon
to the double-pulse excitation process.
We found that the constructive and destructive 
interference between the two overlapped pulses affect the 
excitation efficiency significantly.
On the average, the overlapped double-pulse increases the 
efficiency about two times compared to the distinct double-pulse case.
Our results indicates that we can enhance and control the excitation of silicon by 
the overlapped double pulse.
Also, our result indicates that the excitation efficiency with two pulses 
increases depending on the time-delay due to the decrease of the band gap
 by the lattice temperature.
 
\section*{Aknowledgements}
This research is supported by MEXT Quantum Leap Flagship Program (MEXT Q-LEAP)
under Grant No.  JPMXS0118067246. 
This research is also partially supported by JST-CREST under Grant No. JP-MJCR16N5.
The numerical
calculations are carried out using the computer facilities of the
SGI8600 at Japan Atomic Energy Agency (JAEA).
\bibliography{Refs}


\end{document}